\documentclass[aps,showpacs,twocolumn,pra]{revtex4-1}

\usepackage{graphicx}
\begin{document}
\title{Aharonov-Bohm effect, local field interaction, and Lorentz invariance}


\author{Kicheon Kang}
\email{kicheon.kang@gmail.com}
\affiliation{Department of Physics, Chonnam National University,
 Gwangju 500-757, Republic of Korea}

\begin{abstract}
A field-interaction scheme is introduced for describing the Aharonov-Bohm effect, fully consistent
with the principle of relativity.
Our theory is based on the fact that local field interactions are present
even when a particle moves
only in a field-free region. The interaction Lagrangian between a charge 
and a flux is uniquely constructed from three principles: 
Lorentz covariance, linearity in the interaction strength, and a
correct stationary limit of charge. Our result resolves fundamental questions
raised on the standard interpretation of the Aharonov-Bohm effect,
concerning its duality with the Aharonov-Casher effect
and the equivalence between the
potential and the field-interaction models for describing the
electromagnetic interaction.
Most of all, potential is eliminated in our theory, and all kind of the
force-free Aharonov-Bohm effect is understood in a unified framework of
the Lorentz-covariant local interaction of electromagnetic fields.
\end{abstract}
%

\maketitle

\section{Introduction}
Aharonov-Bohm (AB) effect~\cite{aharonov59,ehrenberg49} has changed our notion 
of electromagnetic field and potential.
It is known as a milestone in our understanding of electromagnetic
interactions, which describes a quantum interference of a charged particle moving
in a region free of electric and magnetic fields.
The general consensus is that the
AB effect demonstrates the invalidity of the classical picture 
of electromagnetism based on the local action of fields.
Aharonov-Casher (AC) effect~\cite{aharonov84}, dual to the AB phenomenon,
shows a phase shift of a magnetic moment (or a fluxon) moving around 
a charged rod. It has been
shown that AC effect is also free of force~\cite{aharonov88,aharonov05},
but standard view draws an apparent distinction between the two phenomena,
although it has been rarely noticed.
In contrast to the AB effect, the incident particle (fluxon) moves under 
a nonvanishing field generated by the
charge in the case of AC effect~\cite{batelaan09}.
Despite this distinction, the two effects are in fact the same
phenomena in two dimension with a charge and a fluxon - 
only their reference frames are different.
Even though the observable phenomena depend only on the 
relative motion of a charge and a fluxon in two dimension, 
a unified picture, fully consistent with the principle of relativity, is lacking.

In this paper, we present a unified theoretical framework which removes 
the reference-frame-dependent ("nonlocal" or "local") description of
the AB and AC effects.  Our framework is based on
a Lorentz-covariant field-interaction between a charge and a localized flux.
The AB effect can be understood in this fully relativistic and local viewpoint.
The AB phase shift is derived from the Lorentz-covariant interaction Lagrangian,
and the force-free nature of the effect is also maintained.
The present study suggests a fundamental change in our notion of the potential:
The AB effect can be described purely in terms of local interaction of fields, and
the role of fields and potentials are unified both in the classical and in the quantum theories.
We believe that our study could initiate
further research to clarify the issue of the nonlocality in electromagnetic action.

The paper is organized as follows.
In Section II, critical questions are addressed concerning the standard, nonlocal
viewpoint of the AB effect. Section III describes our 
local-interaction-based Lorentz-covariant approach to the AB effect.
Conclusion is given in Section IV.

\section{Questions concerning the standard viewpoint of the Aharonov-Bohm
 effect}
First, let us consider the interaction of a point charge, $e$, and a
localized fluxon with its flux value $\Phi$,
in two spatial dimension~\cite{aharonov84,aharonov05} (Figure~1(a)).
The observable phenomena here depend only on the relative motion of
the two entities, in that (1) a phase shift is acquired for one-loop rotation
of one particle moving around the other, and,
(2) the interaction between the particles is free of force in both cases.
In spite of this equivalence, the standard view draws a clear distinction
between the two cases,
namely ``Type I" and ``Type II"~(Figure~1(b))~\cite{batelaan09}. In Type I,
a charge is moving
in a field-free region. The observed AB phase is interpreted as a
pure topological effect where the charge is interacting with the flux only in
a nonlocal way.
In Type II, a neutral particle with a magnetic moment (fluxon, in our case)
 moves under the influence of the electric
field generated by the charge~\cite{aharonov84}. Although the net force applied to the
fluxon vanishes~\cite{aharonov88}, the observed phase shift can be fully
understood in terms of local interaction of the fluxon with the external 
electric field~\cite{peshkin95}.
This distinction of the types depending on the reference
frame should not be inherent, because the AB and the AC effects 
in two dimension are in fact
the same problem in that only a relative motion of the two objects matters.
A single unified description, fully consistent with the principle of relativity,
is desirable for understanding both types of the AB effect.
%


Next, we address the the two different ways of describing 
the electromagnetic interaction between the charge and fluxon.
In fact, equivalence of the two different approaches for describing 
the electromagnetic interaction
energy is well established in classical
electrodynamics (see e.g., Ref.~\onlinecite{griffiths99}).
For a current distribution ($\mathbf{j_1}$)
in the presence of a magnetic field 
$\mathbf{B}_2$~$(=\nabla\times\mathbf{A}_2)$ from an independent
source,  the interaction energy of the two
entities is
given by $\frac{1}{c}\int\mathbf{j}_1\cdot\mathbf{A}_2 d\mathbf{x}$,
coupling of the current and the vector potential ($\mathbf{A}_2$).
Alternatively, it can be expressed as
$\frac{1}{4\pi}\int\mathbf{B}_1\cdot\mathbf{B}_2 d\mathbf{x}$, where
$\mathbf{B}_1$ is the magnetic field generated by the current $\mathbf{j}_1$.
When it is applied to the AB effect, the general consensus is
that only the potential-based approach is valid.
It is hard to see why and how the equivalence is broken
in a quantum mechanical treatment of the interaction. To 
prove that only the potential-based model is
relevant, it would be necessary to show how the equivalence of the
two model is broken. 
In the following, we show that this is not the case, 
and that the equivalence of the two
approaches is preserved also in quantum theory.

\section{Lorentz-invariant field-interaction approach}
The questions raised in Section II are resolved in our theoretical framework
presented here. We develop a
field-interaction scheme which is fully consistent with the
special theory of relativity.
The local field interaction of two particles
(charge and fluxon) in two dimension is treated 
on an equal footing (Figure~1(a)) (The usual picture of a
test particle moving around an external source is inappropriate
here).
For moving charge and fluxon (with their velocities $\dot{\mathbf{r}}$ and
$\dot{\mathbf{R}}$, respectively), there are two sources
of field interactions, (1) between the magnetic field of the fluxon
and that of the moving charge, 
and, (2) between the electric field of the charge and that of the moving
fluxon. 
Each interaction term was treated for a
classical force explanation of the AB~\cite{liebowitz65,boyer73,boyer02}, and
of the Aharonov-Casher (AC)~\cite{boyer87} phase shifts, respectively.
Although the classical force based on the field interactions was refuted both
theoretically~\cite{aharonov88,aharonov05} and experimentally~\cite{caprez07},
we point
out that the field-interaction framework itself is not erroneous.
The main drawback of the previous field-interaction approaches
is that the interaction term depends on the reference frame, 
violating the principle of relativity.
Below, we provide a Lorentz-covariant
field interaction scheme to get rid of this problem.

\subsection{Construction of the Lagrangian}
The Lagrangian, $L$, of the system is given by $L=L_e+L_\Phi+L_{\rm int}$,
where $L_e$, $L_\Phi$, and $L_{\rm int}$ represent a free
charge ($e$), a fluxon ($\Phi$), and their interaction, respectively.
The self field energies diverge for point particles and are therefore
neglected. In any case, the self field energy does not affect our analysis.
The interaction term, $L_{\rm int}=\int {\cal L}_{\rm int} d\mathbf{x}$,
is constructed from the following three first principles:
(1) The Lagrangian density ${\cal L}_{\rm int}$ is invariant
under Lorentz transformation and space inversion, (2) linear in
field strengths, and (3) $L_{\rm int}$ is reduced to $-eV$, the correct
limit of the stationary charge, 
with the electric scalar potential $V$ generated by the moving
fluxon. It is surprising that the interaction Lagrangian is uniquely 
determined from these constraints as
%
%
\begin{equation}
 L_{\rm int} = \frac{1}{8\pi} \int F_{\mu\nu}^{(e)} F^{\mu\nu(\Phi)}
   \,d\mathbf{x} \,,
\label{eq:Lint}
\end{equation}
where $F_{\mu\nu}^{(e)}$ and $F^{\mu\nu(\Phi)}$ are the electromagnetic
field tensors
generated by the charge and the fluxon, respectively. 
Eq.~(\ref{eq:Lint})
can also be written in terms of more familiar electric and
magnetic fields as
\begin{equation}
 L_{\rm int} = \frac{1}{4\pi} \int
    \left( \mathbf{B}^{(e)}\cdot\mathbf{B}^{(\Phi)}
         - \mathbf{E}^{(e)}\cdot\mathbf{E}^{(\Phi)}
    \right) d\mathbf{x} \,,
\label{eq:Lint2}
\end{equation}
where $\mathbf{B}^{(e)}$ ($\mathbf{E}^{(e)}$) and $\mathbf{B}^{(\Phi)}$
($\mathbf{E}^{(\Phi)}$) represent the magnetic (electric) fields
of the charge and the fluxon, respectively.
$L_{\rm int}$ can be simplified by adopting the relations
$\mathbf{B}^{(e)}=\frac{1}{c}\dot{\mathbf{r}}\times\mathbf{E}^{(e)}$ and
$\mathbf{E}^{(\Phi)}=-\frac{1}{c}\dot{\mathbf{R}}\times\mathbf{B}^{(\Phi)}$,
as,
\begin{equation}
 L_{\rm int} = (\dot{\mathbf{r}} - \dot{\mathbf{R}})\cdot \vec{\Pi} \,,
\label{eq:Lint-pi}
\end{equation}
where
\begin{equation}
 \vec{\Pi} = \frac{1}{4\pi c} \int \mathbf{E}^{(e)}\times \mathbf{B}^{(\Phi)}
       d\mathbf{x}
\end{equation}
is the field momentum produced by the two particles.
Note that Eq.~(\ref{eq:Lint-pi}) is equivalent to the standard
interaction Lagrangian
of~\cite{aharonov84,aharonov05}
\begin{equation}
 L_{\rm int} = \frac{e}{c} (\dot{\mathbf{r}} - \dot{\mathbf{R}})\cdot 
 \mathbf{A} \,,
\end{equation}
based on the vector potential $\mathbf{A}$, except that the gauge
dependence is absent in the former (Eq.~(\ref{eq:Lint-pi})).
%

%


In the low speed limit~($|\dot{\mathbf{r}}|, |\dot{\mathbf{R}}| \ll c$), 
the Lagrangian of the system is reduced to
\begin{equation}
 L = \frac{1}{2} m \dot{\mathbf{r}}\cdot\dot{\mathbf{r}}
   + \frac{1}{2} M \dot{\mathbf{R}}\cdot\dot{\mathbf{R}}
   + (\dot{\mathbf{r}} - \dot{\mathbf{R}})\cdot \vec{\Pi} \,,
\label{eq:lagrangian-NR}
\end{equation}
where the field momentum is
\begin{displaymath}
 \vec{\Pi} = \frac{e\Phi}{2\pi c|\mathbf{r}-\mathbf{R}|} \hat{\phi} ,
\end{displaymath}
with $\hat{\phi}$ being !the azimuthal unit vector
of $\mathbf{r}-\mathbf{R}$.
$m$ and $M$ denote the masses of the charge and the fluxon, respectively.
This Lagrangian can also be transformed to a Hamiltonian
\begin{equation}
 H = \frac{(\mathbf{p}-\vec{\Pi})^2}{2m} + \frac{(\mathbf{P}+\vec{\Pi})^2}{2M},
\label{eq:hamiltonian-NR}
\end{equation}
where $\mathbf{p}$ and $\mathbf{P}$ are the canonical momenta of the variables
$\mathbf{r}$ and $\mathbf{R}$, respectively.

\subsection{Understanding the force-free Aharonov-Bohm effect}
Several noticeable features can be found from the Lagrangian of
Eq.~(\ref{eq:lagrangian-NR}) (or from the Hamiltonian of Eq.~(\ref{eq:hamiltonian-NR})).
First, this Lagrangian is equivalent
to the one represented by the
vector potential~\cite{aharonov84,aharonov05},
except that there is no freedom to choose a gauge in Eq.~(\ref{eq:lagrangian-NR}).
The equivalence of the ``potential" and the ``field-interaction" approaches 
is restored.

Second, the Euler-Lagrange equation for the variable $\mathbf{r}$,
\begin{displaymath}
 \frac{d}{dt}\left(\frac{\partial L}{\partial\dot{r}_i}\right) 
   - \left(\frac{\partial L}{\partial {r}_i}\right) = 0 , 
\end{displaymath}
leads to 
\begin{equation}
 m\ddot{\mathbf{r}} =  q(\dot{\mathbf{r}}-\dot{\mathbf{R}})
  \times \mathbf{B}^{(\Phi)} = 0  \,.
\end{equation}
Similarly, for the fluxon's variable $\mathbf{R}$, we get
\begin{equation}
 M\ddot{\mathbf{R}} = -m\ddot{\mathbf{r}} = 0.
\end{equation}
That is, there is no mutual classical force
between the two particles, contrary to the previous classical
explanation of the AB effect based on the field
interactions~\cite{liebowitz65,boyer73,boyer87,boyer02}.
This implies that the classical force claimed in the previous
schemes is an error
which results from ignoring the relativistic invariance.
For instance, if $-\dot{\mathbf{R}}\cdot\Pi$ term (which corresponds to the
electric field interaction) is neglected in
Eq.~(\ref{eq:lagrangian-NR}), the Lagrangian does not satisfy the relativistic
invariance, and we find that $M\ddot{\mathbf{R}}\ne0$ (whereas 
$m\ddot{\mathbf{r}}=0$), violating Newton's
3rd law.
This was the basis of the claim in Ref.~\cite{boyer73}. Apparently,
this force is absent if the Lorentz invariance is taken into account in
the Lagrangian.

Third, on encircling around the fluxon, the charge acquires an AB phase
\begin{equation}
 \phi_{AB} = \frac{1}{\hbar} \oint \vec{\Pi}\cdot d\mathbf{r}
           = \frac{e\Phi}{\hbar c} .
\label{eq:AB-phase}
\end{equation}
The same is true for the fluxon encircling around the charge.
Recall that, in obtaining this result, we have not relied on
the vector potential. Force-free AB phase can be explained in a general way
without the notion of potential, in contrast to the standard nonlocal viewpoint.
%

It is usually believed that the local-interaction model without a potential cannot properly take
account of the force-free AB effect. Part of the reason for this view originates from the reaction force argument claimed in Ref.~\onlinecite{liebowitz65,boyer73,boyer87,boyer02}, which was proven to be incorrect. For example, it was shown~(page 25-26 of Ref.~\onlinecite{peshkin89}) that the erroneous force, derived from the interaction energy 
$\frac{1}{4\pi}\int\mathbf{B}^{(e)}\cdot\mathbf{B}^{(\Phi)}\,d\mathbf{x}$, is removed if the electromotive force (emf) generated by the moving charge is also taken into account in the field energy. We believe that this is one of the main reasons why the local-interaction approach, which is quite natural,
has been widely overlooked. However, the latter contribution (emf) 
is included in our framework in the electric term 
$-\frac{1}{4\pi}\int\mathbf{E}^{(e)}\cdot\mathbf{E}^{(\Phi)}\,d\mathbf{x}$
of the interaction Lagrangian (Eq.~\ref{eq:Lint2}). The erroneous reaction 
force is absent in our local-interaction model. 

Our findings are not limited to the two point particles in two dimension, 
but can be
applied to more realistic case with distributed flux (charge).
Consider, for example, a charge ($e$) and distributed magnetic flux schematically
drawn in Figure~2. The interaction Lagrangian in this case
is given in the same form of Eq.~(\ref{eq:lagrangian-NR})
with the field momentum replaced by
\begin{equation}
 \vec{\Pi} = \frac{e}{2\pi c}
  \int_S \frac{\hat{\phi}B\,d^2\mathbf{R}}{|\mathbf{r}-\mathbf{R}|} \,.
\end{equation}
The integral is over the cross sectional area $S$ of the flux lines,
and
$\hat{\phi}$ is the azimuthal unit vector of $\mathbf{r}-\mathbf{R}$.
The force-free nature of the interaction is maintained. 
Also, it is straightforward to show that the one-loop integral of $\vec{\Pi}$
gives the AB phase of $\phi_{AB} = e\Phi/(\hbar c)$, where $\Phi$ is
the net magnetic flux.

%
\section{Conclusion}
Since the discovery of the AB effect, it has become a common notion that
a framework based on the local action of fields is impossible in
quantum mechanics. 
Here we have provided an alternative, unified framework based on the
Lorentz-covariant field-interaction. It shows that the AB phase shift
can be universally described in terms of the local interaction of fields.
Our result does not reduce the significance of the AB
effect, nor that of its various applications, such as the concept of 
gauge field.
It suggests, however, the following crucial change in our understanding
of the electromagnetic interaction in quantum theory.
First, vector potential is eliminated in our scheme, and thus,
the force-free AB effect can be
explained purely in terms of local field interactions.
This possibility was recently addressed with some specific examples by
Vaidman~\cite{vaidman12}. Our result shows, in a general way,
that there is no physical
effect in the absence of a local overlap of fields.
Second, our study restores the equivalence of the potential-based and the
field-interaction-based schemes in quantum theory of electromagnetic
interactions. 
The equivalence is already present in the Lagrangian of the system, 
and therefore,
there is no reason to discard the field-interaction approach
for treating a quantum mechanical problem.
Third, the AB effect can be described in a unified framework 
independent of the reference frame, without a need for distinction
of its type~\cite{batelaan09}. With this unified viewpoint,
the principle of relativity is fully restored.
Fourth, there is no freedom to choose a gauge for a
potential in our scheme, simply because the Lagrangian is uniquely determined
by the field strengths. This is also closely connected to the
fact that the flux-dependent local phase variation is
uniquely determined in an Aharonov-Bohm loop~\cite{kang12}.
Finally, it is
straightforward to apply the Lorentz-covariant field-interaction framework
to the electric AB effect.
This implies that all kind of the AB effect can be described in a
unified approach.

\section*{Note added in the 5th version}
The section with the title ``Questions about shielding" in the former version
is removed, as it dealt with the shielding classically which
leads to misleading arguments. For more complete analysis on the shielding
of the local interaction of fields, see our later work of 
Ref.~\onlinecite{kang15}.

\section*{Acknowledgments}
%
This work was supported by the National Research Foundation of Korea under
Grant No. 2012R1A1A2003957.



%



\begin{figure}
\includegraphics[width=3.6in]{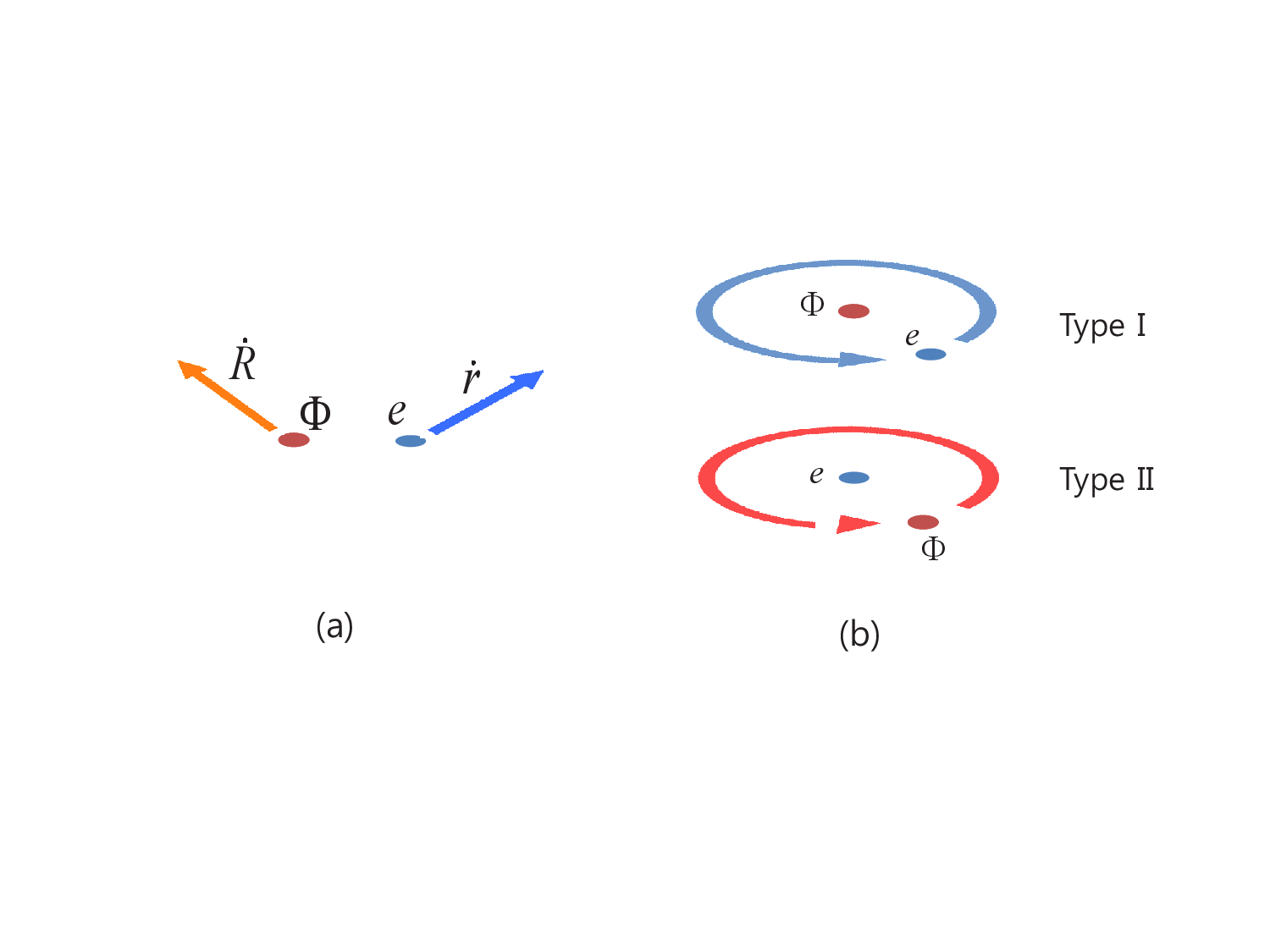} 
\caption{(a) A charge ($e$) and a fluxon ($\Phi$) in two dimension, 
 moving with
 velocities $\mathbf{\dot{r}}$ and $\mathbf{\dot{R}}$, respectively.
(b) Classification of the Aharonov-Bohm effect.
 In ``Type I", a charge is moving in a field-free region, whereas in ``Type II",
 a fluxon moves under the influence of an electric field produced by a charge.
 }
\end{figure}
\begin{figure}
\includegraphics[width=4.5in]{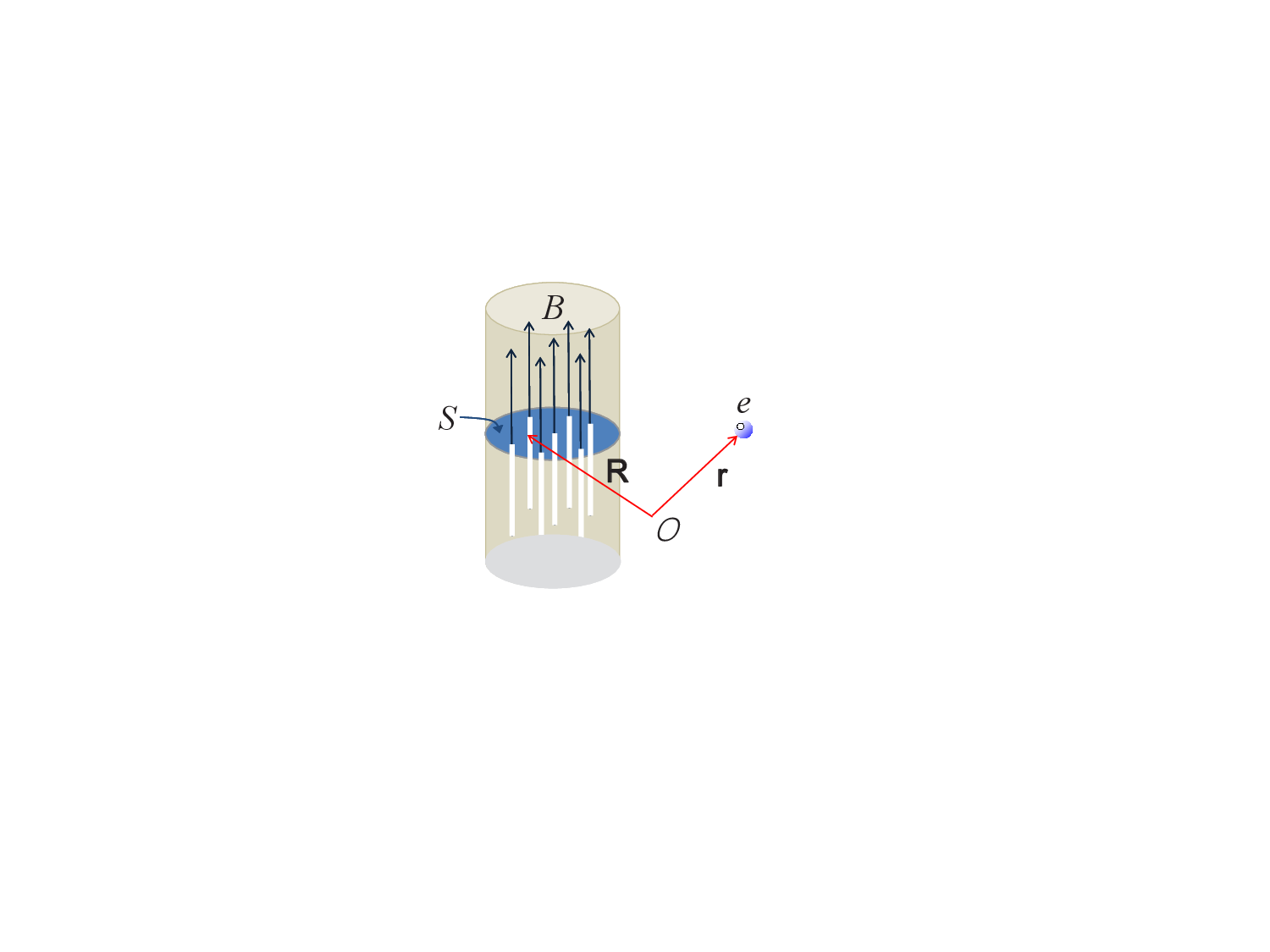} 
\caption{A moving charge in the presence of a distribution of 
 the magnetic field
 $\mathbf{B}$.
 }
\end{figure}
%
%
%



%
\end{document}